\documentclass[letterpaper,12pt]{article}
\pdfoutput=1
\usepackage{jcappub}
\usepackage{epsfig}
\usepackage{amsmath}
\usepackage{subfigure}

\DeclareSymbolFont{AMSa}{U}{msa}{m}{n}
\DeclareSymbolFont{AMSb}{U}{msb}{m}{n}
\let\Box\relax
\DeclareMathSymbol{\Box}{\mathord}{AMSa}{"03}

\newcommand{\be}{\begin{equation}}
\newcommand{\ee}{\end{equation}}
\newcommand{\bea}{\begin{eqnarray}}
\newcommand{\eea}{\end{eqnarray}}

\newcommand{\f}{\frac}
\newcommand{\h}{\hspace*{1mm}}

\newcommand{\ds}{\displaystyle}

\newcommand{\Od}{{\mathcal{O}}}
\newcommand{\Rd}{{\mathcal{R}}}

\title{Bogoliubov Excited States and the Lyth Bound}

\author{Aditya Aravind,}
\author{Dustin Lorshbough,}
\author{and Sonia Paban}
\affiliation{Department of Physics and Texas Cosmology Center\\ The University of Texas at Austin,
TX 78712.}
\emailAdd{Aditya@physics.utexas.edu}
\emailAdd{Lorsh@utexas.edu}
\emailAdd{Paban@zippy.ph.utexas.edu}

\abstract{We show that Bogoliubov excited scalar and tensor modes do not alleviate Planckian evolution during inflation if one assumes that $r$ and the Bogoliubov coefficients are approximately scale invariant.  We constrain the excitation parameter for the scalar fluctuations, $\beta$, and tensor perturbations, $\tilde{\beta}$, by requiring that there be at least three decades of scale invariance in the scalar and tensor power spectrum.  For the scalar fluctuations this is motivated by the observed nearly scale invariant scalar power spectrum.  For the tensor fluctuations this assumption may be shown to be valid or invalid by future experiments.}

\arxivnumber{1403.6216}

\begin{document}
\maketitle
\flushbottom
\section{Introduction}
The recent measurement by BICEP2 of the primordial B-mode polarization \cite{Ade:2014xna} indicate that naively the inflaton field traverses Planckian field values during inflation according to the Lyth bound \cite{Lyth:1996im}.  There have been attempts to alter the expression for the tensor-to-scalar ratio in a way such that this conclusion no longer holds \cite{Dufaux:2007pt,Cook:2011hg,Senatore:2011sp,Barnaby:2011qe,Barnaby:2012xt,Carney:2012pk,Biagetti:2013kwa,Lello:2013awa,Collins:2014yua}.  In particular, there have been investigations into whether excited spectator fields or modified scalar/tensor mode functions may appreciably alleviate the need for Planckian evolution.\\

Any loop correction to the scalar and/or tensor power spectrum will be suppressed by powers of the Planck mass and therefore are not promising for generating an appreciable effect in general.  Additionally, any loop corrections to the scalar power spectrum must be small so as not to disrupt the agreement between standard inflationary theory and the observed scalar power spectrum.  Therefore, rather than relying on loop corrections one promising avenue for generating an appreciable modification to the tensor-to-scalar ratio is through modifications of the fluctuation mode functions.  We will show that for scale invariant Bogoliubov parameters\footnote{There have previously been investigations into the consequences of one particular form of tensor excitation scale dependence \cite{Ashoorioon:2013eia,Ashoorioon:2014nta}.  These studies use different arguments than are presented here.  Also, see \cite{Sriramkumar:2004pj} for a discussion on the relationship between trans-Planckian physics and excited initial states.}, excited Bogoliubov states do not appreciably alleviate the need for Planckian evolution.\\

We will denote the reduced Planck mass by $M_P^2=(8\h\pi\h G_N)^{-1}$ and physical momenta by $p=(k/a)$.  For simplicity we will assume that the tensor-to-scalar ratio $r$ is approximately scale independent over the three to four decades of the observed scalar power spectrum, this assumption may be shown to be invalid by future experiments.

\section{Case A: Vacuum $\Rd$ and Vacuum $\gamma$}\label{sec:CaseA}
We begin by reviewing the standard derivation of the Lyth Bound.  The scalar and tensor fluctuations may be expanded in classical mode functions as follows \cite{Maldacena:2002vr}
\begin{equation}\label{eq:flds}
\hat{\Rd}=\int\h\f{d^3k}{(2\pi)^3}\h\hat{\Rd}_k\h e^{i\vec{k}\cdot\vec{x}}, \h\h\h\hat{\gamma}_{ij}=\int\h\f{d^3k}{(2\pi)^3}\h\underset{s=\pm}{\Sigma}\h\epsilon_{k,ij}^s\h\hat{\gamma}_{\vec{k}}^s\h e^{i\vec{k}\cdot\vec{x}},
\end{equation}
\begin{equation}
\hat{\Rd}_{\vec{k}}=\Rd_k\h \hat{a}_{\vec{k}}^\dagger+\Rd_k^*\h \hat{a}_{-\vec{k}},\h\h\h \hat{\gamma}_{\vec{k}}^s=\gamma_{k}^s\h \hat{b}_{\vec{k}}^{s\dagger}+\gamma_k^{*s}\h \hat{b}_{-\vec{k}}^s,
\end{equation}
\begin{equation}
[\hat{a}_k,\hat{a}_{k'}^\dagger]=(2\h\pi)^3\delta^3\left(\vec{k}-\vec{k}'\right),\h\h\h[\hat{b}_k^s,\hat{b}_{k'}^{s'\dagger}]=(2\h\pi)^3\delta^3\left(\vec{k}-\vec{k}'\right)\delta^{s,s'}.
\end{equation}
Here we have introduced the polarization tensor $\epsilon_{k,ij}^s$ such that\footnote{For a graviton moving in the $\hat{z}$ direction we find $\Lambda_{ijlm}(\hat{k})\equiv\underset{s}{\Sigma}\epsilon_{k,ij}^s\epsilon_{k,lm}^s=\delta_{il}\delta_{jm}+\delta_{im}\delta_{jl}-\delta_{ij}\delta_{lm}+\delta_{ij}\hat{k}_l\hat{k}_m+\delta_{lm}\hat{k}_i\hat{k}_j-\delta_{il}\hat{k}_j\hat{k}_m-\delta_{im}\hat{k}_j\hat{k}_l-\delta_{jl}\hat{k}_i\hat{k}_m-\delta_{jm}\hat{k}_i\hat{k}_l+\hat{k}_i\hat{k}_j\hat{k}_l\hat{k}_m$.  The normalization is chosen so that $\Lambda_{1111}(k)=1$ since the non-zero polarization tensor elements are given by  $\epsilon_{k,11}^{s=\pm}=-\epsilon_{k,22}^{s=\pm}=\mp i\epsilon_{k,12}^{s=\pm}=\mp i\epsilon_{k,21}^{s=\pm}=\f{1}{\sqrt{2}}$. By noting that $\Lambda_{ijij}(\hat{k})=4$ one may immediately verify that $\epsilon_{k,ij}^s\epsilon_{k,lm}^{s'}=2\h\delta^{ss'}$ since in that case $\Lambda_{ijij}(\hat{k})=\f{1}{2}\Lambda_{ijlm}\Lambda_{ijlm}$ \cite{Weinberg:2008zzc}.} $\epsilon_{k,ij}^s\epsilon_{k,ij}^{s'}=2\h\delta^{ss'}$, $\epsilon_{k,ii}^s=0$ and $k^i\h\epsilon_{k,ij}^s=0$.  There are two possible helicity values for $s$.  The second order actions, classical evolution equations and corresponding mode functions for the fluctuations are given by
\begin{equation}\label{eq:action}
S_\Rd=-M_P^2\h\epsilon\h\int\h d^4x\h \sqrt{-\bar{g}}\h \bar{g}^{\mu\nu}\h \partial_\mu\hat{\Rd}\h \partial_\nu\hat{\Rd},\h\h\h S_\gamma=-\f{M_P^2}{8}\h\int\h d^4x\h \sqrt{-\bar{g}}\h \bar{g}^{\mu\nu}                                                                                                                                  \h \partial_\mu\hat{\gamma}_{ij}\h \partial_\nu\hat{\gamma}_{ij},
\end{equation}
\begin{equation}\label{eq:eom_A}
\ddot{\Rd}_k+3\h H\h \dot{\Rd}_k+\f{k^2}{a^2}\Rd_k=0,\h\h\h \ddot{\gamma}_{k}^s+3\h H\h \dot{\gamma}_{k}^s+\f{k^2}{a^2}\gamma_k^s=0,
\end{equation}
\begin{equation}\label{eq:mode_A}
\Rd_{k,A}=\f{1}{\sqrt{2\h\epsilon}}\f{H}{M_P\sqrt{2\h k^3}}\h\left(1+i\f{k}{a\h H}\right)e^{-ik/aH},\h\h\h \gamma_{k,A}^s=\sqrt{2}\f{H}{M_P\h\sqrt{2\h k^3}}\h\left(1+i\f{k}{a\h H}\right)e^{-ik/aH}.
\end{equation}
Here $\bar{g}$ denotes the background value of the metric.  We have introduced the slow roll parameter $\epsilon$ which, after introducing the number of e-folds $dN=H\h dt$, we will define as
\begin{equation}\label{eq:epsilon_A}
\epsilon=\f{1}{2}\f{\dot{\phi}^2}{M_P^2\h H^2}=\f{1}{2\h M_P^2}\left(\f{d\phi}{dN}\right)^2.
\end{equation}
The corresponding late time power spectra are given by \cite{Ade:2013uln}
\begin{equation}\label{eq:PS_0}
\langle\hat{\Rd}_k\h\hat{\Rd}_{k'}\rangle=(2\h\pi)^3\h\delta^3\left(\vec{k}+\vec{k}'\right)P_\Rd,\h\h\h 4\langle\hat{\gamma}_k^s\h\hat{\gamma}_{k'}^{s}\rangle=(2\h\pi)^3\h\delta^3\left(\vec{k}+\vec{k}'\right)P_\gamma,
\end{equation}
\begin{equation}\label{eq:PS_A}
\Delta_\Rd^2=\f{k^3}{2\pi^2}P_\Rd=\f{1}{8\pi^2\epsilon}\f{H^2}{M_P^2},\h\h\h\Delta_\gamma^2=\f{k^3}{2\pi^2}P_\gamma=\f{2}{\pi^2}\f{H^2}{M_P^2}.
\end{equation}
The tensor-to-scalar ratio is given by
\begin{equation}\label{eq:r_A}
r_A=\f{\Delta_{\gamma}^2}{\Delta_{\Rd}^2}=16\h\epsilon.
\end{equation}
The Lyth bound in terms of the usual Planck mass is given by combining (\ref{eq:epsilon_A}) and (\ref{eq:r_A})
\begin{equation}\label{eq:Lyth_A}
\left(\f{\Delta \phi}{M_P}\right)_A=\int_0^N\h dN'\h\sqrt{2\h\epsilon}=\f{1}{\sqrt{8}}\h\int_0^N\h dN'\h\sqrt{r}.
\end{equation}
Using $r_A=0.2$ \cite{Ade:2014xna} we find that over the range of say $\Delta N=6.9$ ($1<l<10^3$) e-folds, the inflaton field evolves by 1.09 $M_P$.  Therefore even within the timespan that the modes comprising the CMB exited the horizon, the background inflaton field evolves on Planckian scales. 

\section{Case B: Excited $\Rd$ and Excited $\gamma$}
Now we allow for an excited mode functions that are of the form of a Bogoliubov transformation
\begin{equation}\label{eq:mode_B}
\Rd_{k,B}=\alpha_k\h \Rd_{k,A}+\beta_k\h\Rd_{k,A}^*,\hspace*{4mm}\gamma_{k,B}^s=\tilde{\alpha}_{k}^s\h\gamma_{k,A}^s+\tilde{\beta}_{k}^s\h\gamma_{k,A}^{s*}=\tilde{\alpha}_{k}\h\gamma_{k,A}^s+\tilde{\beta}_{k}\h\gamma_{k,A}^{s*}.
\end{equation}
We will assume for simplicity that $\tilde{\alpha}_{k}^s$ and $\tilde{\beta}_{k}^s$ are the same for both polarizations.  The new power spectra are given by
\begin{equation}\label{eq:PS_B}
\Delta_\Rd^2=\f{k^3}{2\pi^2}P_\Rd=\f{1}{8\pi^2\epsilon}\f{H^2}{M_P^2}|\alpha_k+\beta_k|^2,\hspace*{4mm}\Delta_\gamma^2=\f{k^3}{2\pi^2}P_\gamma=\f{2}{\pi^2}\f{H^2}{M_P^2}|\tilde{\alpha}_k+\tilde{\beta}_k|^2.
\end{equation}
The new tensor-to-scalar ratio is given by
\begin{equation}\label{eq:r_B}
r_B=16\h\epsilon\h\f{|\tilde{\alpha}_k+\tilde{\beta}_k|^2}{|\alpha_k+\beta_k|^2}.
\end{equation}
Combining (\ref{eq:Lyth_A}) and (\ref{eq:r_B}) we obtain the new Lyth bound
\begin{equation}\label{eq:Lyth_B}
\left(\f{\Delta \phi}{M_P}\right)_B=\f{1}{\sqrt{8}}\h\int_0^N\h dN'\h\sqrt{r_A}\f{|\alpha_k+\beta_k|}{|\tilde{\alpha}_k+\tilde{\beta}_k|}\approx\f{|\alpha+\beta|}{|\tilde{\alpha}+\tilde{\beta}|}\left(\f{\Delta \phi}{M_P}\right)_A.
\end{equation}
In the last equality we assumed that $\alpha_k$ and $\beta_k$ are approximately constant over the range of integration due to the observed nearly scale invariant scalar power spectrum\footnote{Note that this conclusion holds even if one reconstructs the power spectrum from Planck measurements alone \cite{Aslanyan:2014mqa} or the combination of Planck and BICEP2 measurements \cite{Abazajian:2014tqa}}.  We additionally have assumed that the tensor excitation parameters are scale invariant as well.  We will discuss this in greater detail in the next section and show that $0.98\lesssim|\alpha+\beta|\lesssim1.02$ and $0.98\lesssim|\tilde{\alpha}+\tilde{\beta}|^{-1}\lesssim1.02$, therefore Planckian field evolution is still present in this case.  Specifically, for the case of three decades of scale invariant power spectra we find
\begin{equation}\label{eq:ratio_bound}
0.96\lesssim\f{|\alpha+\beta|}{|\tilde{\alpha}+\tilde{\beta}|}\lesssim 1.04
\end{equation}
Therefore exciting the tensor and/or scalar modes in an approximately scale independent manner will not allow one to alleviate Planckian evolution.
\subsection{Constraints on $|\beta|$}\label{sec:AppA}
This section summarizes some of the results from \cite{Aravind:2013lra,Flauger:2013hra}.  The modes comprising the cosmic microwave background lay between some low energy and high energy cutoff.
\begin{equation}\label{eq:p_range_beta}
p_{IR}<p_{\text{obs}}<p_{UV}.
\end{equation}
We observe between three and four decades ($10^n$) of scale invariant power spectrum modes in the cosmic microwave background.  In order  for the modes observed today to exit horizon during inflation, the low momentum modes must be inside the horizon when CMB modes first begin leaving the horizon, $p_{IR}>H$.  In order not to disrupt slow-roll inflation, the high energy momentum modes must obey the backreaction bound, namely the fluctuation energy density should be less than the background energy density, $\langle\rho_\Rd\rangle\ll H^2 M_P^2$.  Moreover, in order to not disrupt the rate of change of the Hubble parameter we require that $\langle\rho_\Rd\rangle\ll \epsilon H^2 M_P^2$.\footnote{This follows from $\epsilon\h H^2=-\dot{H}$ and the Einstein equation $(2M_P^2)\dot{H}=-(\rho+P)$, where $P_{\Rd,\gamma}=\rho_{\Rd,\gamma}/3$.  The form of the action given in (\ref{eq:action}) requires that $-\dot{H}/H^2=\dot{\phi}^2/2M_P^2H^2$, which is only true if this constraint is satisfied.}  Combining these constraints allows us to write
\begin{equation}\label{eq:ineq_beta}
10^n=\f{p_{UV}}{p_{IR}}\ll\f{p_{\text{Back-Reaction}}}{H}.
\end{equation}
The energy density of scalar fluctuations is given by
\begin{equation}\label{eq:rho_beta}
\begin{array}{lll}
\left\langle \rho_{\Rd} \right\rangle&=&\ds{M_P^2\h\epsilon\int_0^{k_{\text{UV}}}\f{d^3k}{(2\pi)^3}\left[|\dot{\Rd}_{k,B}|^2-|\dot{\Rd}_{k,A}|^2+\left(\f{k}{a}\right)^2\left(|\Rd_{k,B}|^2-|\Rd_{k,A}|^2\right)\right]}\\
&=&\ds{\f{|\beta|^2}{8\pi^2}\left[p_{\text{
UV}}^4+\Od\left(H^2p_{\text{UV}}^2\right)\right].}\end{array}
\end{equation}
We have assumed that $|\beta_k|\approx|\beta|$ is approximately scale invariant over the modes of interest.  This is justified by the scale invariance of the scalar power spectrum.  For other models found in the literature, such as an exponential dependence \cite{Holman:2007na,Ganc:2011dy}, the exponential is nearly constant and nearly unity for the observed modes as discussed in \cite{Aravind:2013lra}.  The back-reaction constraint therefore implies
\begin{equation}\label{eq:bound_BR}
p_{\text{Back-Reaction}}=(8\h\pi^2\h \epsilon)^{1/4}\sqrt{\f{H\h M_P}{|\beta|}}.
\end{equation}
Substituting (\ref{eq:bound_BR}) into (\ref{eq:ineq_beta}) and solving for $|\beta|$ we find
\begin{equation}\label{eq:H_ineq_beta}
|\beta|<10^{-2\h n}(8\h\pi^2\h \epsilon)^{1/2}\left(\f{M_P}{H}\right)=10^{-2n}\left(\Delta_\Rd^2\right)^{-1/2}|\alpha+\beta|.
\end{equation}
Note that the scalar power spectrum (\ref{eq:PS_B}) has an observed amplitude of $\Delta_\Rd^2\approx2\times 10^{-9}$ \cite{Ade:2013zuv}.  Combining this with (\ref{eq:r_B}) we obtain
\begin{equation}\label{eq:master_beta}
|\beta|\lesssim 2|\alpha+\beta|\times\left\{\begin{array}{ll}10^{-2}&,n=3\\10^{-4}&,n=4
\end{array}\right..
\end{equation}
Using $|\alpha+\beta|\leq|\alpha|+|\beta|=|\beta|+\sqrt{1+|\beta|^2}$ we find that even in the conservative limit $|\beta|\lesssim 0.02$.  Since $|\alpha|\approx1$ for this small a $|\beta|$, we find that $0.98\lesssim|\alpha+\beta|\lesssim1.02$.

\subsection{Constraints on $|\tilde{\beta}|$}\label{sec:AppB}
The argument presented here is nearly identical to that used in section (\ref{sec:AppA}).  The energy density of the tensor fluctuations is given by \cite{Kundu:2013gha}
\begin{equation}\label{eq:rho_tensor}
\begin{array}{lll}
\left\langle \rho_{\gamma} \right\rangle&=&\ds{\underset{s=\pm}{\ds{\Sigma}}\h\f{2\h M_P^2}{8}\int_0^{k_{\text{UV}}}\f{d^3k}{(2\pi)^3}\left[|\dot{\gamma}_{k,B}^s|^2-|\dot{\gamma}_{k,A}^s|^2+\left(\f{k}{a}\right)^2\left(|\gamma_{k,B}^s|^2-|\gamma_{k,A}^s|^2\right)\right]}\\
&=&\ds{\f{|\tilde{\beta}|^2}{4\h\pi^2}\left[p_{\text{
UV}}^4+\Od\left(H^2p_{\text{UV}}^2\right)\right].}\end{array}
\end{equation}
We have assumed that the Bogoliubov coefficients are approximately scale independent for the tensor modes, or at least are as scale independent as the scalar coefficients were in the previous case.  This is a simplifying assumption that observations may either justify by finding an approximately scale invariant tensor spectrum or invalidate by finding the contrary.\\

The analog of (\ref{eq:H_ineq_beta}) for the tensor case is (taking n=3)
\begin{equation}\label{eq:H_ineq_tensor}
|\tilde{\beta}|<(2\h\pi)\h 10^{-2\h n}\sqrt{\epsilon}\left(\f{M_P}{H}\right)=10^{-2\h n}\f{|\alpha+\beta|}{(2\h \Delta_\Rd^2)^{1/2}}\lesssim0.02.
\end{equation}
Finally noting that $|\tilde{\alpha}|^2=1+|\tilde{\beta}|^2$ we find $0.98\lesssim|\tilde{\alpha}+\tilde{\beta}|^{-1}\lesssim1.02$.

\section{Conclusions}
We have shown that modifying the scalar and/or tensor fluctuation mode functions to have a non-zero Bogoliubov excitation parameter does not allow one to escape having Planckian field evolution according to the Lyth bound if one assumes at least three decades of scale independence for the tensor-to-scalar ratio ($r$) and the Bogoliubov excitation parameters ($\beta,\tilde{\beta}$).  Further measurements may show that the scale dependence in the tensor spectrum is stronger than in the scalar spectrum implying the scale dependence of $\tilde{\beta}$ is non-negligible.  An interesting direction for future studies would be to find out what degree of scale dependence is necessary to prevent the presence of Planckian field evolution and finding what physical mechanism generates such scale dependence.

\section*{Acknowledgments}
We would like to thank Sandipan Kundu for helpful discussions.  This material is based upon work supported by the National Science Foundation under Grant Number PHY-1316033.
%

\end{document}